# An Efficient Bilinear Pairing-Free Certificateless Two-Party Authenticated Key Agreement Protocol in the eCK Model

Yong-Jin Kim *[1], Yong-Min Kim[1], Yong-Jin Choe[1] and Hyong-Chol O[1]

[1] *Faculty of Mathematics,* **KimIlSung** *University, Pyongyang, DPR of Korea*

* Corresponding author, Email: ohyongchol@yahoo.com

**ABSTRACT**: Recent studyoncertificateless authenticated key agreement focuses on bilinear pairing-freecertificateless authenticated key agreement protocol. Yet it has got limitations in the aspect of computational amount.So it is important to reduce the number of the scalar multiplication over elliptic curve group in bilinear pairing-free protocols. This paper proposed a new bilinear pairing-freecertificatelesstwo-party authenticated key agreement protocol, providing more efficiency among related work and proofunder the random oracle model.

## 1. Introduction

Introduction of mobile devices including mobile phones, laptops and etc. has made human life more comfortable. However, it has also brought urgent problems to be solved in the communication through *insecure* channels.

Al-Riyami and Paterson [1]introduced a new certificateless public key cryptography (CLPKC).CLPKCavoidedthe certificate management problems in the traditional public key cryptosystem (PKC) and the inherent key escrow problem in the identity-based(ID-based) public key cryptosystem[2]. The CLPKCis intermediate between the traditional PKC and ID-based cryptosystem. In a certificateless cryptosystem,a user's private key is not generated by the key generationcenter(KGC) alone. Instead, it consists of a partialprivate key generated by the KGC and some secret value chosen by the user. Thus, the KGC isunable to obtain the user's private key in such a way that the key escrow problem can be solved.Intuitively, CLPKC has nice features borrowed from both ID-based cryptography and traditionalPKC. It alleviates the key escrow problem in ID-based cryptography and at the same time reducesthe cost and simplifies the use of the technology when compared with traditional PKC.

Following the pioneering work due to Al-Riyami and Paterson, several certificatelesstwo-party authenticated key agreement (CTAKA) protocols[3–6] have been proposed. All theabove CTAKA protocolsmay be practical, but they are from bilinear pairings. The relative computation cost of apairing is approximately 20 times higher than that of the scalar multiplication over elliptic curvegroup [7].

Therefore, CTAKA protocolswithout bilinear pairings would be more appealing in termsof efficiency.The severalbilinear pairing-free CTAKA protocols have beenproposed [8, 9, 10, 11, 12, 19].Yang et al [10] pointed out that neither Geng et al. [8]'s protocol nor Hou et al. [9]'s protocol is secure. They proposed an improvedCTAKA protocol to improve the security. He et al [11] also proposed abilinear pairing-free CTAKA protocol but it is vulnerable to thetype 1 adversary [12, 20]. Bellare et al. [13] is the first to propose a formal security model for authentication and key distribution. Since then, there have been several extensions[14, 15, 16] given to the model. Among them, the modified Bellare-Rogaway (mBR) model [13] and the Canetti-Krawczyk (CK) model [16]are regarded as promising ones. In 2007,LaMacchia et al [17] presented a considerably strong security model—the extended Canetti-Krawczyk (eCK) model. The eCK model captures many desirable security properties including key-compromise impersonation (KCI) resilience, weak perfect forward secrecy (wPFS) and ephemeral secrets reveal resistance etc. while the originalCK model does notcover KCI attacks.

From the description of the eCK model for CTAKA protocolin the following section 2.6, one can know that the previous bilinear pairing-free CTAKA protocols [8, 9, 11,12] are not secure in the eCK model. The protocols [10, 19]are provably secure in the eCK model. However, the user in protocols [10, 19] needs 9 and 5 elliptic curve scalar multiplications to finish the key agreement.Then it is necessary to design efficient bilinear pairing-free CTAKA protocol, which is provably secure in the eCK model.

In this paper, with the purpose of reducing the amount of computation, we shall propose an efficient bilinear pairing-free CTAKA scheme and prove our protocol is provably secure under the eCK model.

The remainder of this paper is organized follows. The section 2 gives some preliminaries. Our new protocol is given in the section 3. The security analysis of the proposed protocol is presented in the section 4. In the section 5, we compare our scheme with previous protocols. Finally, in the section 6 we provide some conclusions.

## 2.    Preliminaries

### 2.1 Notations

For convenience, some notations used in this paper are described as follows.
- $p, q$: two large prime numbers
- $F_p$: prime field
- $E / F_p$: an elliptic curve defined over a prime field$F_p$
- $G$: the cyclic additive group composed of the point on $E / F_p$
- $P$: a generator of $G$
- $H_1(\cdot)$: a secure one-way hash function
- $H_2(\cdot)$: a secure one-way hash function
- $ID_i$: the identity of user $i$
- $s$:the master private key of the KGC
- $P_{KGC}$: the public key of the KGC
- $r_i$: the KGC's ephemeralprivate key
- $D_i$: the user $i$'s partial private key
- $x_i$: the user $i$'s secret value
- $S_i$: the user $i$'s private key
- $P_i$: the user $i$'s public key
- $t_i$:the user $i$'s ephemeral private key
- $T_i$: the user $i$'s ephemeral public key
- $k$: the security parameter
- $params$: the system parameter

### 2.2 Process in configuring a new key agreement protocol

The process in configuring the new protocol and proving its security comes in following steps.
   a) Provide a computational hardness problem.
   b) Provide the security model of the new protocol.

c) Provide the new protocol.
d) Provide a provable security in the security model.
  ➢ Provide a formal definition of the goals of the protocol.
  ➢ Reduce the advantage of the adversary against the protocol to solving of the computational hardness problems.
  ➢ Prove the security of the protocol in the way of deriving the contradiction of the above assumption.

**2.3 Computational hardness problem**

The following problems defined over $G$ are assumed to be intractable within polynomial time.
  ➢ **Discrete Logarithm (DL) problem**: For $a \in_R Z_q^*$ and $P$ the generator of $G$, given $(P, a \cdot P)$, compute $a$.
  ➢ **Computational Diffie-Hellman (CDH) problem**: For $a, b, c \in_R Z_q^*$ and $P$ the generator of $G$, given $(P, a \cdot P, b \cdot P)$, compute $ab \cdot P$.
  ➢ **Decisional Diffie-Hellman (DDH) problem**: For $a, b, c \in_R Z_q^*$ and $P$ the generator of $G$, given $(P, a \cdot P, b \cdot P, c \cdot P)$, decide whether $ab \cdot P = c \cdot P$ or not.
  ➢ **Gap Diffie-Hellman (GDH) problem**: For $a, b, c \in_R Z_q^*$ and $P$ the generator of $G$, given $(P, a \cdot P, b \cdot P, c \cdot P)$, compute $ab \cdot P$, along with full access to a decision oracle DDH.

The GDH assumption states that the probability of any polynomial-time algorithm to solve the GDH problem is negligible.

**2.4 Desirable Security Properties for two-party authenticated key agreement protocols**

The following security properties are commonly required for two-party authenticated key agreement protocols:
  ➢ Resistance to Basic Impersonation Attacks (BIS): An adversary who does not know the private key of party $A$ should not be able to impersonate $A$.
  ➢ Resistance to Unknown Key-Share (UKS) Attacks: It should be impossible to coerce $A$ into thinking it is sharing a key with $B$, when it is actually sharing a key with another (honest) user $C$ (and $C$ correctly thinks the key is shared with $A$).
  ➢ Known Key Security (KKS): Each run of a key agreement protocol between two parties $A$ and $B$ should produce a unique session key. A protocol should not become insecure if the adversary has learned some of the session keys.
  ➢ Resistance to Key-Compromise Impersonation (KCI) Attacks: If the private key of a user $A$ is compromised, the attacker should not be able to impersonate another user $B$ to $A$.
  ➢ Weak Perfect Forward Secrecy (WPFS): An attacker who knows the private keys of all parties, but is not actively involved in choosing ephemeral keys during the sessions of interest, should not be able to determine previously established session keys.
  ➢ Resistance to Disclosure of Ephemeral Secrets (DES): The protocol should be resistant to the disclosure of ephemeral secrets. The disclosure of an ephemeral secret should not compromise the security of other sessions.
  ➢ KGC Forward Secrecy (KGC-FS): Certificateless protocols usually require that the KGC should be unable to compute previously established session keys even if it knows all publicly available information.
  ➢ Resistance to Leakage of Ephemeral Secrets to KGC (KGC-LES): In certificateless protocols, the KGC that has learned the ephemeral secrets of any session should not be able to compute the session key.

## 2.5  CTAKA protocol

ACTAKA protocol consists of six polynomial-time algorithms [2-5]: *Setup, Partial-Private-Key-Extract, Set-Secret-Value, Set-Private-Key, Set-Public-Key and Key-Agreement.* These algorithms are respectively defined as follows.

- $Setup$:
  - ✓ Input: $k$
  - ✓ Output: $params, s$
- $Partial - Private - Key - Extract$:
  - ✓ Input: $params, s, ID_i$
  - ✓ Output: $D_i$
- $Set - Secret - Value$:
  - ✓ Input: $params, ID_i$
  - ✓ Output: $x_i$
- $Set - Private - Key$:
  - ✓ Input: $params, ID_i, x_i$
  - ✓ Output: $S_i$
- $Set - Public - Key$:
  - ✓ Input: $params, ID_i, x_i$
  - ✓ Output: $P_i$
- $Key - Agreement$:
  - ✓ Input: $(ID_i, S_i, P_i), (ID_j, S_j, P_j)$
  - ✓ Output: $SK_{ij} = SK_{ji} = SK$

## 2.6  Security model for CTAKA protocols

In CTAKA, as defined in [1], there are two types of adversaries with different capabilities.

**Definition 1:** We say that an adversary is an *outside attacker* if the adversary does not have the KGC's master secret key. We assume an outside attacker is able to replace public keys of users. The outside attacker is called the *type 1 adversary* $\mathcal{A}_1$.

**Definition 2**: We say that an adversary is an *inside attacker* if the adversary has access to the KGC's master secret key. We assume an inside attacker cannot replace public keys of users. The inside attacker is called the *type 2 adversary* $\mathcal{A}_2$.

Let $U = \{U_1, U_2, \cdots, U_n\}$ be a set of parties. The protocol may be run between any two of these parties. For each party there exists an identity. There is a KGC that issues identity based partial private keys to the parties through secure channel. Additionally, the parties generate their own secret values and corresponding certificateless public keys. The adversary is in control of the network over which protocol messages are exchanged. $\prod_{i,j}^{t}$ represents the $t$-th protocol session which runs at party $i$ with intended partner party $j$. A session $\prod_{i,j}^{t}$ enters an *accepted* state when it computes a session key $SK_{ij}^{t}$. Note that a session may terminate without ever entering into an accepted state. The information of whether a session has terminated with acceptance or without acceptance is assumed to be public. The session $\prod_{i,j}^{t}$ is assigned a partner ID $pid = (ID_i, ID_j)$. The session ID $sid$ of $\prod_{i,j}^{t}$ at party $i$ is the transcript of the messages exchanged with party $j$ during the session. Two sessions $\prod_{i,j}^{t}$ and $\prod_{j,i}^{l}$ are considered matching if they have the same $pid$ and $sid$. The eCK model in the CLPKC setting is defined by the following game between a *challenger* τ and an adversary $\mathcal{A} \in \{\mathcal{A}_1, \mathcal{A}_2\}$. In the model, $\mathcal{A}$ is modeled by a probabilistic polynomial-time Turing machine (PPT). All communications go through the adversary $\mathcal{A}$. Participants only respond to the queries by $\mathcal{A}$ and do not communicate directly among themselves. $\mathcal{A}$ can relay, modify, delay, interleave or delete all the message flows in the system. $\mathcal{A}$ may ask a polynomial number of the following queries as follows.

The game runs in two phases.

During the phase of the game, the adversary $\mathcal{A}$ is allowed to issue the following queries in any order:

$Create(i)$: This allows $\mathcal{A}$ to ask the $\tau$ to set up a new participant $i$ with identity $ID_i$. On receiving such a query, the $\tau$ generates the private/public key pair for $i$.

$RevealPartialPrivateKey(i)$: the $\tau$ responds with $i$'s partial private key.

$RevealSecreteValue(i)$: the $\tau$ responds with $i$'s secret value $x_i$ that corresponds to its certificateless public key. If the $\tau$ has been asked the replace public key query before, it responds with $\perp$.

$ReplacePublicKey(i, pk)$: Party's certificateless public key is replaced with $pk$ chosen by the adversary. Party $i$ will use the new public key for all communication and computation.

$RevealEphemeralKey(\prod_{i,j}^{t})$: the $\tau$ responds with the ephemeral secret used in session

$RevealMasterKey$: The adversary is given access to the master secret key.

$RevealSessionKey(\prod_{i,j}^{t})$: If the session has not been accepted, it returns $\perp$, otherwise it reveals the accepted session key.

$Send(\prod_{i,j}^{t}, m)$: If the session $\prod_{i,j}^{t}$ does not exist, it will be created as initiator at party $i$ if $m = \lambda$, or as a responder at party $j$, otherwise. If theparticipating parties have not been initiated before, the respective private and public keys are created. Upon receiving the message $m$, the protocol is executed. After party $i$ has sent and received the last set of messages specified by the protocol, it outputs a decision indicating accepting or rejecting the session. In the case of one-round protocols, party $i$ behaves as follows:

$m = \lambda$: Party $i$ generates an ephemeral value and responds with an outgoing message only and a decision indicating acceptance or rejection of the session.

$m \neq \lambda$: If party $i$ is a responder, it generates an ephemeral value for thesession and responds with an outgoing message and a decision indicating acceptance or rejection of the session. In this work, we require $i \neq j$, i.e. a party will not run a session with itself.

Once the adversary $\mathcal{A}$ decides that the first phase is over, it starts the second phase by choosing a fresh session $\prod_{i,j}^{t}$ and issuing a $Test(\prod_{i,j}^{t})$ query, where the fresh session and $Test$ query are defined as follows.

**Definition 3** (Fresh session): A session $\prod_{i,j}^{t}$ is fresh if
- ➢ $\prod_{i,j}^{t}$ has accepted;
- ➢ $\prod_{i,j}^{t}$ is unopened(not being issued the session key reveal query);
- ➢ The session state at neither party participating in this session is fully corrupted;
- ➢ There is no opened session $\prod_{j,i}^{l}$ which has a matching conversation to $\prod_{i,j}^{t}$.

$Test(\prod_{i,j}^{t})$: At some point, $\mathcal{A}$ may choose one of the oracles, say $\prod_{i,j}^{t}$, to aska single $Test$ query. This oracle must be fresh. If $b = 0$, the adversary is given the session key, otherwise it randomly samples a session key from the distribution of valid session keys and returns it to the adversary.

After the $Test(\prod_{i,j}^{t})$ query, $\mathcal{A}$ can continue to query except that the test session $\prod_{i,j}^{t}$ should remain fresh. We emphasize here that partial corruption is allowed as this is a benefit of our security model. Additionally, $ReplacePublicKey$ queries may be issued to any party after the test session has been completed.

At the end of the game, $\mathcal{A}$ must output a guess bit $b'$. $\mathcal{A}$ wins if and only if $b' = b$. $\mathcal{A}$'s advantage in winning the above game is defined as:

$$Adv_{\mathcal{A}}(k) = \left|\Pr(b' = b) - \frac{1}{2}\right|.$$

**Definition 4**. A CTAKA protocol is said to be secure if:
> In the presence of a benign adversary on $\prod_{i,j}^{t}$ and $\prod_{j,i}^{u}$, both oracles always agree on the same session key, and this key is distributed uniformly at random.
> For any adversary $\mathcal{A} \in \{\mathcal{A}_1, \mathcal{A}_2\}$, $Adv_{\mathcal{A}}(k)$ is negligible.

## 3. Our protocol

Our protocol also consists of six polynomial-time algorithms. They are described as follows.

**Setup:**

The KGC randomly picks $s \in Z_q^*$ as a master private key and sets its public key $P_{KGC} = s \cdot P$.

For a security parameter $k$, KGC selects two cryptographic secure hash functions $H_1: \{0,1\}^k \to Z_q^*$, $H_2: \{0,1\}^* \times \{0,1\}^* \times G \times G \times G \times G \to Z_q^*$.

The KGC publishes $params = \{F_p, E/F_p, G, P, P_{KGC}, H_1, H_2\}$ as system parameters and secretly keeps the master key $s$.

**Partial-Private-Key-Extract:**

The KGC chooses at random $r_i \in Z_q^*$, computes $R_i = r_i \cdot P$, $h_i = H_1(ID_i, R_i)$, $s_i = r_i + h_i s$ mod $q$ and issues user's partial private key $D_i = (s_i, R_i)$ to the users with identity $ID_i$ through secret channel.

He can validate her private key by checking whether the equation $s_i \cdot P = R_i + h_i \cdot P_{KGC}$ holds. The private key is valid if the equation holds and vice versa.

**Set-Secret-Value:**

The user with identity $ID_i$ picks randomly $x_i \in Z_q^*$ and sets $x_i$ as his secret value.

**Set-Private-Key:**

The user with identity $ID_i$ takes the pair $S_i = (x_i, s_i)$ as its private key.

**Set-Public-Key:**

The user with identity $ID_i$ takes $P_i = x_i \cdot P$ as its public key.

**Key-Agreement:**

Each user $A, B$:
> User Keys and Message Exchange
>> ✓ A
>>> • chooses a random number $t_A \in Z_q^*$ and computes $T_A = t_A \cdot P$.
>>> • sends $M_1 = (ID_A, R_A, T_A)$ to $B$.
>> ✓ B
>>> • chooses a random number $t_B \in Z_q^*$ and computes $T_B = t_B \cdot P$.
>>> • sends $M_2 = (ID_B, R_B, T_B)$ to $A$.
> Key Computation

For two parties $A$ and $B$:
>> ✓ Upon receiving $M_2$ from $B$, $A$ computes
>>> • $u_A = x_A + s_A + t_A$

- $W_B = R_B + H_1(ID_B, R_B) \cdot P_{KGC}$
- $K_{AB}^1 = u_A \cdot (P_B + W_B)$
- $K_{AB}^2 = u_A \cdot (T_B + W_B)$
- $SK_{AB} = H_2(ID_A||ID_B||T_A||T_B||K_{AB}^1||K_{AB}^2)$

✓ Upon receiving $M_1$ from $A$, $B$ computes
- $W_A = P_A + R_A + H_1(ID_A, R_A) \cdot P_{KGC} + T_A$
- $K_{BA}^1 = (x_B + s_B) \cdot W_A$
- $K_{BA}^2 = (t_B + s_B) \cdot W_A$
- $SK_{BA} = H_2(ID_A||ID_B||T_A||T_B||K_{BA}^1||K_{BA}^2)$

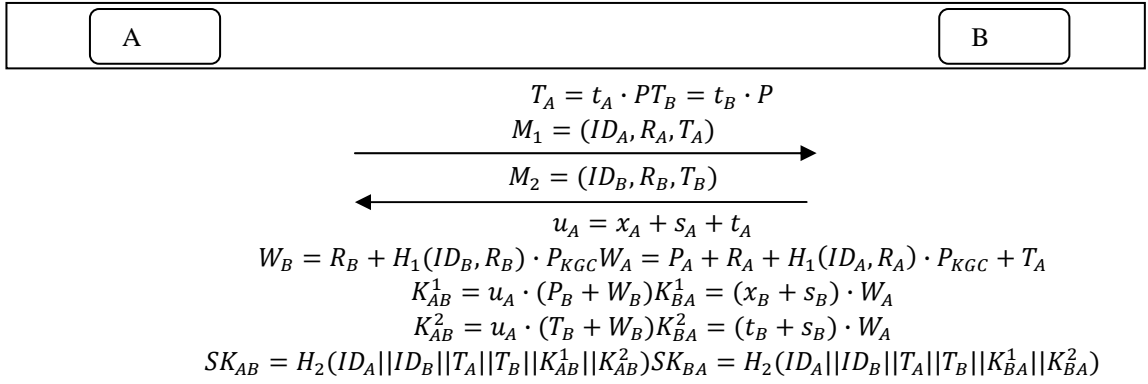

Fig.1. Key agreement of our protocol

The shared secrets agree because:
$$T_A = t_A \cdot P, \; P_A = x_A \cdot P, \; s_A \cdot P = R_A + H_1(ID_A, R_A) \cdot P_{KGC}$$
$$T_B = t_B \cdot P, \; P_B = x_B \cdot P, \; s_B \cdot P = R_B + H_1(ID_B, R_B) \cdot P_{KGC}$$
$$u_A = x_A + s_A + t_A$$
$$W_B = R_B + H_1(ID_B, R_B) \cdot P_{KGC}$$
$$K_{AB}^1 = u_A \cdot (P_B + W_B) = u_A \cdot (P_B + R_B + H_1(ID_B, R_B) \cdot P_{KGC})$$
$$= (x_A + s_A + t_A)(x_B + s_B) \cdot P$$
$$=(x_B + s_B) \cdot (P_A + R_A + H_1(ID_A, R_A) \cdot P_{KGC} + T_A) = (x_B + s_B) \cdot W_A = K_{BA}^1$$

and
$$K_{AB}^2 = u_A \cdot (T_B + W_B) = u_A \cdot (T_B + R_B + H_1(ID_B, R_B) \cdot P_{KGC})$$
$$= (x_A + s_A + t_A)(t_B + s_B) \cdot P$$
$$=(t_B + s_B) \cdot (P_A + R_A + H_1(ID_A, R_A) \cdot P_{KGC} + T_A) = (t_B + s_B) \cdot W_A = K_{BA}^2$$

Thus, the agreed session keys for A and B can be computed as:
$SK_{AB} = H_2(ID_A||ID_B||T_A||T_B||K_{AB}^1||K_{AB}^2) = H_2(ID_A||ID_B||T_A||T_B||K_{BA}^1||K_{BA}^2) = SK_{BA} = SK$

The correctness of the protocol is proved.

## 4. Security Analysis

To prove the security of our protocol in the random oracle model, we assume $H_1$ and $H_2$ as two random oracles [18]. For security analysis we use proof method of the reference [19].

**Theorem 1.** *If it is difficult to solve the GDH problem, we can neglect the advantage of a type 1 adversary against our protocol.*

**Proof:** Suppose that there is a type 1 adversary $\mathcal{A}_1$ who can win the game defined in the section 2 with a non-negligible advantage $Adv_{\mathcal{A}_1}(k)$ in polynomial-time $t$. Then we will show how to use the ability of $\mathcal{A}_1$ to construct an algorithm τ to solve the GDH problem. The τ first chooses $P_0 \in_R G$ as $P_{KGC}$, selects the system parameter $params = \{F_p, E/F_p, G, P, P_{KGC}, H_1, H_2\}$ and sends $params$ to $\mathcal{A}_1$.

Let $n_0$ be the maximum number of sessions that any one party may have. Assume that the adversary $\mathcal{A}_1$ activates at most $n_1$ distinctive honest parties and activates at most $n_2$ distinctive hash queries.

Assume also that $Adv_{\mathcal{A}_1}(k)$ is non-negligible. Since $H_1$, $H_2$ are modeled as a random oracle, after the adversary issues the test query, it has only three possible ways to distinguish the tested session key from a random string:

**CASE 1**.*Guessing attack*: $\mathcal{A}_1$ guesses correctly the session key.

**CASE 2**.*Key-replication attack*: The adversary $\mathcal{A}_1$ forces a non-matching session to have the same session key with the test session. In this case, the adversary $\mathcal{A}_1$ can simply learn the session key by querying the non-matching session.

**CASE 3**.*Forging attack*: Assume that $\prod_{I,J}^{T}$ is the test session. At some point in its run, the adversary $\mathcal{A}_1$ queries $H_2$ in the test session owned by $I$ communicating with $J$. Clearly, $\mathcal{A}_1$ computes the values of $K_{IJ}^1$ and $K_{IJ}^2$ in this case.

Since $H_2$ is a random oracle, the probability of guessing the output of $H_2$ is $(1/2^k)$, which is negligible.

The input to the key derivation function $H_2$ includes all information that can uniquely identify the matching sessions. Since two non-matching sessions cannot have the same identities and the same ephemeral public keys and $H_2$ is modeled as a random oracle, the success probability of the *Key-replication attack* is also negligible. Thus *Guessing attack* and Key-replication attack can be ruled out, and the rest of the proof is mainly devoted to the analysis of the *Forging attack*. As the attack that the adversary $\mathcal{A}_1$ mounts is the forging attack, $\mathcal{A}_1$ cannot get an advantage in winning the game against the protocol unless it queries the $H_2$ oracle on the session key. To relate the advantage of the adversary $\mathcal{A}_1$ against our protocol to the GDH assumption, we use a classical reduction approach. In the following, a challenger τ is interested to use the adversary $\mathcal{A}_1$ to turn $\mathcal{A}_1$'s advantage indistinguishing the tested session key from a random string into an advantage in solving the GDH problem. Let $Adv_{GDH}(k)$ be the advantage that the challenger τ gets in solving the GDH problem given the security parameter $k$. To solve the GDH problem using $\mathcal{A}_1$, the τ is given a GDH challenge $U = u \cdot P, V = v \cdot P$ and an oracle $DDH(*,*,*)$, where $u, v \in Z_q^*$, and the τ's task is to compute $GDH(U,V) = uv \cdot P$. The τ simulates the game outlined in the section 2. During the game, the τ has to answer all queries of the adversary $\mathcal{A}_1$. Before the game starts, the τ tries to guess the test session and the strategy that the adversary $\mathcal{A}_1$ will adopt. $\mathcal{A}_1$ randomly selects $T \in \{1,2,\cdots n_0\}$ and two indexes $I, J \in \{1,\ldots,n\}: I \neq J$, which represent the $I$-th and the $J$-th distinct honest party that the adversary initially chooses. And then the τ determines the test session $\prod_{I,J}^{T}$, which is correct with probability larger than $1/n_0 n_1^2$.

Let $\prod_{I,J}^{T}$ be the matching session of $\prod_{J,I}^{L}$. The following two sub-cases should be considered.

**CASE 3.1**: The test session has a matching session owned by another honest party.

**CASE 3.2**: No honest party owns a matching session to the test session.

▲ The Analysis of **CASE 3.1**

The strongest adversary is allowed to corrupt at most two out of three secrets for each party as follows:

- $x_I$: the user $I$'s secret value;
- $s_I$: the part of the user $I$'s partial private key;
- $t_I$: the user $I$'s ephemeral private key.

This gives to the adversary nine possibilities, called *strategies* (Lippold et al.[5]), to break the protocol. In Lippold et al. [5] $\mathcal{A}$ has 9 stratgies as follows:

**CASE 3.1.1**: $\mathcal{A}$ may neither $x_I$ nor $x_J$.
**CASE 3.1.2**: $\mathcal{A}$ may neither $t_I$ nor $t_J$.
**CASE 3.1.3**: $\mathcal{A}$ may neither $x_I$ nor $t_J$.
**CASE 3.1.4**: $\mathcal{A}$ may neither $t_I$ nor $x_J$.
**CASE 3.1.5**: $\mathcal{A}$ may neither $s_I$ nor $x_J$.
**CASE 3.1.6**: $\mathcal{A}$ may neither $x_I$ nor $s_J$.
**CASE 3.1.7**: $\mathcal{A}$ may neither $s_I$ nor $t_J$.
**CASE 3.1.8**: $\mathcal{A}$ may neither $t_I$ nor $s_J$.
**CASE 3.1.9**: $\mathcal{A}$ may neither $s_I$ nor $s_J$.

Weshould provide the proof of CASE 3.1.2, CASE3.1.7, CASE 3.1.8 andCASE 3.1.9 as $\mathcal{A}$ is type 1 adversary $\mathcal{A}_1$.

The Analysis of CASE 3.1.2

The $\tau$ answers $\mathcal{A}_1$'s queries as follows.

- $Create(i)$: the $\tau$ maintains an initially empty list $L_c$: $(ID_i, s_i, R_i, x_i, P_i)$. The $\tau$ chooses three random numbers $s_i, h_i, x_i \in Z_q^*$, computes $R_i = s_i \cdot P - h_i \cdot P_0$, $P_i = x_i \cdot P$, sets $H_1(ID_i, R_i) \leftarrow h_i$ and stores $(ID_i, s_i, R_i, x_i, P_i)$ and $(ID_i, R_i, h_i)$ in $L_c$ and $L_{H_1}$ seperately.

- $H_1(ID_i, R_i)$: the $\tau$ maintains an initially empty list $L_{H_1}$: $H_1(ID_i, R_i)$. If $(ID_i, R_i)$ is on the $L_{H_1}$, the $\tau$ returns $h_i$. Otherwise, the $\tau$ randomly pick $h_i \in Z_q^*$s, stores $(ID_i, R_i, h_i)$ in $L_{H_1}$ and returns $h_i$.

- $H_2(ID_i, ID_j, T_i, T_j, Z_1, Z_2, SK)$: the $\tau$ maintains an initially empty list $L_{H_2}$: $(ID_i, ID_j, T_i, T_j, Z_1, Z_2, SK)$. If $(ID_i, ID_j, T_i, T_j, *, *, *)$ is in the list $L_{H_2}$, the $\tau$ responds $SK$. Otherwise, the $\tau$ responds to these queries in the following way:

  ✓ The $\tau$ looks the list $L_s$ for entry $(ID_i, ID_j, T_i, T_j, *)$. If the $\tau$ finds the entry, he computes
    ◆ $ID_i = ID_I$
    $$W_j = R_j + H_1(ID_j, R_j) \cdot P_0$$
    $$\overline{Z_1} = Z_1 - (x_i + s_i) \cdot (P_j + W_j) - s_j \cdot T_i$$
    $$\overline{Z_2} = Z_2 - (x_i + s_i) \cdot (T_j + W_j) - s_j \cdot T_i$$
    ◆ $ID_i = ID_J$
    $$W_j = P_j + R_j + H_1(ID_j, R_j) \cdot P_0 + T_j$$
    $$\overline{Z_1} = Z_1$$
    $$\overline{Z_2} = Z_2 - s_i \cdot W_j - s_i \cdot T_j$$

  ✓ Then the $\tau$ checks whether $\overline{Z_d}, d = 1,2$ is correct by checking whether the oracle DDH$(*,*,*)$ outputs 1 when the tuple $(*,*,\overline{Z_d})$, $d = 1,2$, i.e. $(T_i, T_j, \overline{Z_1})$ is inputted. If $Z_1, Z_2$ are correct, the $\tau$ stores the tuple $(ID_i, ID_j, T_i, T_j, Z_1, Z_2, SK)$ into $L_{H_2}$, where the value $SK$ comes from $L_s$. Otherwise, the $\tau$ chooses a random number $SK \in \{0,1\}^k$ and stores the tuple $(ID_i, ID_j, T_i, T_j, Z_1, Z_2, SK)$ into $L_{H_2}$.

- $RevealPartialPrivateKey(i)$: the $\tau$ looks up the list $L_c$ and returns the corresponding $D_i$ to the adversary $\mathcal{A}_1$.

- $RevealSecretValue(i)$: the $\tau$ looks up the list $L_C$ for entry $(ID_i,*,*,*,*)$. If the $\tau$ find the entry, he returns $x_i$. Otherwise, the $\tau$ carries out the query $Create(i)$ and returns the corresponding $x_i$.
- $ReplacePublicKey(i, pk)$: the $\tau$ looks up the list $L_C$ for entry $(ID_i,*,*,*,*)$. If the $\tau$ find the entry, he replaces $x_i$ and $P_i$ with $x'_i, P'_i$, where $pk = P'_i$, $P'_i = x'_i \cdot P$. Otherwise, the $\tau$ carries out the query $Create(i)$ and replaces $x_i$ and $P_i$ with $x'_i, P'_i$.
- $RevealEphemeralKey(\prod_{i,j}^t)$
  - ✓ If $\prod_{i,j}^t = \prod_{I,J}^T$ and $\prod_{i,j}^t = \prod_{J,I}^L$, then the $\tau$ stops the simulation.
  - ✓ Otherwise the $\tau$ returns the stored ephemeral private key to $\mathcal{A}_1$.
- $RevealMasterKey$ : the $\tau$ stops the simulation.
- $RevealSessionKey(\prod_{i,j}^t)$:
  - ✓ If $\prod_{i,j}^t = \prod_{I,J}^T$ or $\prod_{i,j}^t = \prod_{J,I}^L$, then the $\tau$ stops the simulation.
  - ✓ Otherwise the $\tau$ returns the session key to $\mathcal{A}_1$.
- $Send(\prod_{i,j}^t, m)$: the $\tau$ maintains an initially empty list $L_s$: $(ID_i, ID_j, T_i, T_j, SK)$.
  - ✓ If $\prod_{i,j}^t = \prod_{I,J}^T$, then the $\tau$ returns $T_i = U$ to $\mathcal{A}_1$.
  - ✓ If $\prod_{i,j}^t = \prod_{J,I}^L$, then the $\tau$ returns $T_j = V$ to $\mathcal{A}_1$.
  - ✓ Otherwise, the $\tau$ replies according to the specification of the protocol.
- $Test(\prod_{i,j}^t)$: the $\tau$ answers $\mathcal{A}_1$'s queries as follows.
  - ✓ If $\prod_{i,j}^t \neq \prod_{I,J}^T$, the $\tau$ stops the simulation.
  - ✓ Otherwise, the $\tau$ generates a random number $SK \in \{0,1\}^k$ and returns it to $\mathcal{A}_1$.

As the attack that adversary $\mathcal{A}_1$ mounts the forging attack, if $\mathcal{A}_1$ succeeds, it must have queried oracle $H_2$ on the form as follows.

◆ $ID_i = ID_I$

$$u_i = x_i + s_i + t_i$$
$$W_j = R_j + H_1(ID_j, R_j) \cdot P_0$$
$$Z_1 = u_i \cdot (P_j + W_j)$$
$$Z_2 = u_i \cdot (V + W_j)$$

◆ $ID_i = ID_J$

$$W_j = P_j + R_j + H_1(ID_j, R_j) \cdot P_0 + U$$
$$Z_1 = (x_i + s_i) \cdot W_j$$
$$Z_2 = (t_i + s_i) \cdot W_j.$$

Here $T_i = U$ is the outgoing message of test session by the simulator and $T_j = V$ is the incoming message from the adversary $\mathcal{A}_1$. To solve $GDH(U,V)$, for all entries in $L_{H_2}$, the $\tau$ randomly picks one entry with the probability $\frac{1}{n_2}$ and the $\tau$ proceeds with following steps: The $\tau$ computes

◆ $ID_i = ID_I$
$$\overline{Z_2} = Z_2 - (x_i + s_i) \cdot (V + W_j) - s_j \cdot U = GDH(U,V)$$

◆ $ID_i = ID_J$
$$W_j = P_j + R_j + H_1(ID_j, R_j) \cdot P_0 + U$$
$$\overline{Z_2} = Z_2 - s_i \cdot W_j - x_j \cdot U = GDH(U,V)$$ and returns $\overline{Z_2}$ as the solution to $GDH(U,V)$.

The advantage of the $\tau$ solving GDH problem satisfies

$$Adv_\tau^{GDH}(k) \geq \frac{1}{9n_0 n_1^2 n_2} Adv_{\mathcal{A}_1}(k).$$

Then $Adv_\tau^{GDH}(k)$ is non-negligible since we assume that $Adv_{\mathcal{A}_1}(k)$ is non-negligible. This contradicts the GDH assumption.

CASE3.1.7, CASE 3.1.8 and CASE 3.1.9 can be solved by the similar method.

▲ The Analysis of **CASE 3.2**

Through the definition of the freshness, the following two cases should be considered.

**CASE 3.2.1**: At some point, the static private key owned by the party $I$ has been revealed by the adversary $\mathcal{A}_1$ (Note that in this case, according to the freshness definition, $\mathcal{A}_1$ is not permitted to reveal ephemeral private key of the test session).

This case can be solved by the same method in the CASE 3.1.8.

**CASE 3.2.2**: The static private key owned by the party $I$ has never been revealed by the adversary $\mathcal{A}_1$. (Note that in this case, according to the freshness definition, $\mathcal{A}_1$ may reveal party $I$'s ephemeral private key in the test session.)

This case can be solved by the same method in the CASE 3.1.7.

If the adversary $\mathcal{A}_1$ succeeds with non-negligible probability in any case above, we can also solve the GDH problem with non-negligible probability, which contradicts the assumed security of GDH problem. So we can conclude that our scheme bases its security on GDH problem.  ∎

**Theorem 2.** *If it is difficult to solve the GDH problem, we can neglect the advantage of a type 2 adversary against our protocol.*

**Proof:** Suppose that there is a type 2 adversary $\mathcal{A}_2$ who can win the game defined in the section 2 with a non-negligible advantage $Adv_{\mathcal{A}_2}(k)$ in polynomial-time $t$.

▲ The Analysis of **CASE 3.1**

Through the definition of the freshness, we should provide the proof of CASE 3.1.1, CASE3.1.2, CASE3.1.3 and CASE3.1.4 as $\mathcal{A}$ is type 2 adversary $\mathcal{A}_2$.

The Analysis of CASE 3.1.1

A challenger $\tau$ answers $\mathcal{A}_2$'s queries as follows.

➤ $Create(i)$: the $\tau$ maintains an initially empty list $L_c$: $(ID_i, s_i, R_i, x_i, P_i)$.
  If $ID_i = ID_I$, the $\tau$ chooses two random numbers $r_i, h_i \in Z_q^*$, computes $R_i = r_i \cdot P, P_i = U$, $s_i = r_i + h_i s \bmod q$ sets $H_1(ID_i, R_i) \leftarrow h_i$ and stores $(ID_i, s_i, R_i, \perp, P_i)$ and $(ID_i, R_i, h_i)$ in $L_c$ and $L_{H_1}$ seperately. Otherwise, if $ID_i = ID_J$, the $\tau$ chooses two random numbers $r_i, h_i \in Z_q^*$, computes $R_i = r_i \cdot P, P_i = V$, $s_i = r_i + h_i s \bmod q$, sets $H_1(ID_i, R_i) \leftarrow h_i$ and stores $(ID_i, s_i, R_i, \perp, P_i)$ and $(ID_i, R_i, h_i)$ in $L_c$ and $L_{H_1}$ seperately. Otherwise, the $\tau$ chooses three random numbers $s_i, h_i, x_i \in Z_q^*$, computes $R_i = r_i \cdot P, P_i = x_i \cdot P$, $s_i = r_i + h_i s \bmod q$, sets $H_1(ID_i, R_i) \leftarrow h_i$ and stores $(ID_i, s_i, R_i, x_i, P_i)$ and $(ID_i, R_i, h_i)$ in $L_c$ and $L_{H_1}$ seperately.

➤ $H_1(ID_i, R_i)$: the $\tau$ maintains an initially empty list $L_{H_1}: H_1(ID_i, R_i)$.
  If $(ID_i, R_i)$ is in the list $L_{H_1}$, the $\tau$ returns $h_i$. Otherwise, the $\tau$ randomly picks $h_i \in Z_q^*$, stores $(ID_i, R_i, h_i)$ in $L_{H_1}$ and returns $h_i$.

➤ $H_2(ID_i, ID_j, T_i, T_j, Z_1, Z_2, SK)$: the $\tau$ maintains an initially empty list $L_{H_2}: (ID_i, ID_j, T_i, T_j, Z_1, Z_2, SK)$. If $(ID_i, ID_j, T_i, T_j, *, *, *)$ is in the list $L_{H_2}$, the $\tau$ responds $SK$. Otherwise, the $\tau$ responds to these queries in the following way:
  ✓ The $\tau$ looks the list $L_s$ for entry $(ID_i, ID_j, T_i, T_j, *)$. If $\tau$ finds the entry, he computes
    ◆ $ID_i = ID_I$

$$W_j = R_j + H_1(ID_j, R_j) \cdot P_0$$
$$\overline{Z_1} = Z_1 - (s_i + t_i) \cdot (P_j + W_j) - s_j \cdot P_i$$
$$\overline{Z_2} = Z_2 - (s_i + t_i) \cdot (T_j + W_j) - t_j \cdot P_i$$

◆ $ID_i = ID_J$

$$W_j = P_j + R_j + H_1(ID_j, R_j) \cdot P_0 + T_j$$
$$\overline{Z_1} = Z_1 - s_i \cdot W_j - (s_j + t_j) \cdot P_i$$
$$\overline{Z_2} = Z_1$$

✓ Then the τ checks whether $\overline{Z_d}$, $d = 1,2$ is correct by checking whether the oracle $DDH(*,*,*)$ outputs 1 when the tuple $(*,*, \overline{Z_d})$, i.e. $(P_i, P_j, \overline{Z_1})$ is inputted. If $Z_1, Z_2$ are correct, the τ stores the tuple $(ID_i, ID_j, T_i, T_j, Z_1, Z_2, SK)$ into $L_{H_2}$, where the value SK comes from $L_s$. Otherwise, the τ chooses a random number $SK \in \{0,1\}^k$ and stores the tuple $(ID_i, ID_j, T_i, T_j, Z_1, Z_2, SK)$ into $L_{H_2}$.

➢ $RevealPartialPrivateKey(i)$ : the τ looks up the list $L_C$ and returns the corresponding $D_i$ to the adversary $\mathcal{A}_2$.

➢ $RevealSecreteValue(i)$: the τ looks up the list $L_C$ for entry $(ID_i, *, *, *, *)$.
  ✓ If $ID_i = ID_I$ or $ID_i = ID_J$. The τ stops the simulation.
  ✓ Otherwise, If the τ find the entry, he returns $x_i$. Otherwise, the τ carries out the query $Create(i)$ and returns the corresponding $x_i$.

➢ $ReplacePublicKey(i, pk)$: the τ stops the simulation.

➢ $RevealEphemeralKey(\prod_{i,j}^t )$
  ✓ If $\prod_{i,j}^t \neq \prod_{I,J}^T$ and $\prod_{i,j}^t \neq \prod_{J,I}^L$, then the τ stops the simulation.
  ✓ Otherwise the τ returns the stored ephemeral private key to $\mathcal{A}_2$.

➢ $RevealMasterKey$ : the τ returns the master key $\mathcal{A}_2$.

➢ $RevealSessionKey(\prod_{i,j}^t )$:
  ✓ If $\prod_{i,j}^t = \prod_{I,J}^T$ or $\prod_{i,j}^t = \prod_{J,I}^L$, then the τ stops the simulation.
  ✓ Otherwise the τ returns the session key to $\mathcal{A}_2$.

➢ $Send(\prod_{i,j}^t, m)$: the τ maintains an initial empty list $L_s$: $(ID_i, ID_j, T_i, T_j, SK)$.
  ✓ If $\prod_{i,j}^t = \prod_{I,J}^T$, then the τ returns $P_i = U$ to $\mathcal{A}_2$.
  ✓ If $\prod_{i,j}^t = \prod_{J,I}^L$, then the τ returns $P_j = V$ to $\mathcal{A}_2$.
  ✓ Otherwise, the τ replies according to the specification of the protocol.

➢ $Test(\prod_{i,j}^t )$: the τ answers $\mathcal{A}_2$'s queries as follows.
  ✓ If $\prod_{i,j}^t \neq \prod_{I,J}^T$, the τ stops the simulation.
  ✓ Otherwise, the τ generates a random number $SK \in \{0,1\}^k$ and returns it to $\mathcal{A}_2$.

As the attack that adversary $\mathcal{A}_2$ mounts the forging attack, if $\mathcal{A}_2$ succeeds, it must have queried oracle $H_2$ on the form as follows.

◆ $ID_i = ID_I$

$$u_i = x_i + s_i + t_i$$
$$W_j = R_j + H_1(ID_j, R_j) \cdot P_0$$
$$Z_1 = u_i \cdot (V + W_j)$$
$$Z_2 = u_i \cdot (T_j + W_j)$$

◆ $ID_i = ID_J$

$$W_j = U + R_j + H_1(ID_j, R_j) \cdot P_0 + T_j$$
$$Z_1 = (x_i + s_i) \cdot W_j$$

$$Z_2 = (t_i + s_i) \cdot W_j.$$

Here $P_i = U$ is the outgoing message of test session by the simulator and $P_j = V$ is the incoming message from the adversary $\mathcal{A}_2$.

Tosolve $GDH(U,V)$, for all entries in $L_{H_2}$ randomly picks one entry with the probability.

The advantage of the $\tau$ solving GDH problem with the advantage $\frac{1}{n_2}$ and the $\tau$ proceeds with following steps: the $\tau$ computes

◆ $ID_i = ID_I$
$$\overline{Z_1} = Z_1 - (s_i + t_i) \cdot (V + W_j) - s_j \cdot U$$
◆ $ID_i = ID_I$
$$W_j = U + R_j + H_1(ID_j, R_j) \cdot P_0 + T_j$$
$$\overline{Z_1} = Z_1 - s_i \cdot W_j - (s_j + t_j) \cdot V = GDH(U,V) \text{ and returns } \overline{Z_1} \text{ as the solution to } GDH(U,V).$$

The advantage of the $\tau$ solving GDH problem satisfies
$$Adv_\tau^{GDH}(k) \geq \frac{1}{9n_0 n_1^2 n_2} Adv_{\mathcal{A}_2}(k).$$

Then $Adv_\tau^{GDH}(k)$ is non-negligible since we assume that $Adv_{\mathcal{A}_2}(k)$ is non-negligible. This contradicts the GDH assumption.

The CASE 3.1.2, CASE 3.1.3 and CASE 3.1.4 can be solved by the similar method.

▲ The Analysis of CASE 3.2

Through the definition of the freshness, the following two cases should be considered.

**CASE 3.2.1**: At some point, the secret value owned by the party *I* has been revealed by the adversary $\mathcal{A}_2$. (Note that in this case, according to the freshness definition, $\mathcal{A}_2$ is not permitted to reveal ephemeral private key of the test session.)

This case can be solved through the same method in the CASE 3.1.4.

**CASE 3.2.2**: The secret value owned by the party *I* has never been revealed by the adversary $\mathcal{A}_2$. (Note that in this case, according to the freshness definition, $\mathcal{A}_2$ may reveal party *I* 's ephemeral private key in the test session.)

This case can be solved through the same method in the CASE 3.1.3.

If the adversary $\mathcal{A}_2$ succeeds with non-negligible probability in any case above, we can also solve the GDH problem with non-negligible probability, which contradicts the assumed security of GDH problem. So we can conclude that our scheme bases its security on GDH problem. ∎

**Proposition1.** *If two sessions are matching, both of them will be accepted and will get the same session key which is distributed uniformly at random in thesession key sample space.*

**Proof:** From the correction analysis of our protocol in Section 3, we know if two oracles are matching, then both of them are accepted and have the same session key. The session keys are distributed uniformly since $r_A, r_B \in Z_q^*$ are selected uniformly during the protocol execution. ∎

From the above three theorems and proposition, we can get the following corollary.

**Corollary1.** *Our protocol is a secure CTAKA protocol in the eCK model under the GDH assumption.*

## 5. Comparison with previous protocols

For the convenience of evaluating the computational cost, we define some notations as follows.

$T_{mul}$: The time of executing a scalar multiplication operation ofpoint.

$T_{add}$: The time of executing an addition operation of point.

$T_{inv}$: The time of executing a modular invasion operation.

$T_h$: The time of executing a one-way hash function.

We will compare the efficiency of our protocol with six CTAKA protocols without pairings i.e. Geng et al.'s protocol [8], Hou et al.'s protocol [9], Yang et al.'s protocol [10] and He et al.'s protocols [11,12, 19]. The table 1 shows the comparison between bilinear pairing-free CTAKA protocols in terms of efficiency, security model and underlying hardness assumptions.

Since the scalar multiplication operation of point is more complicated than the addition operation of points, modular invasion operation and the hash function operation, then our protocol has better performance than Geng et al.'s protocol [8], Hou et al.'s protocol [9] and Heetal.'s protocol [11, 12, 19]. Moreover, Geng et al.'s protocol [8], Hou et al.'s protocol [9] , Yang et al.'s protocol [10] and He et al.'s protocol [11] are not secure against type adversary. Then our protocol has advantage in both the performance and the security over previous ones. It is well known that the eCK model is much superior to the mBR model. Then Yang et al.'s protocol [10], He et al.'s protocol [19] and our protocol has advantage in security to Geng et al.'s protocol [8] , Hou et al.'s protocol [9] and He et al.'s protocol [11, 12]. From the table 1, we know our protocol has much better performance than previous ones. We conclude that our protocol is more suitable for practical applications.

Table 1: Comparisons among different protocols

| Protocol | Computational amount | Security model | Assumption | Message exchange |
|---|---|---|---|---|
| Geng(2009),[8] | $7T_{mul} + 2T_h$ | mBR | GDH | 2 |
| Hou(2009),[9] | $6T_{mul} + 2T_h$ | mBR | GDH | 2 |
| Yang(2011),[10] | $9T_{mul} + 2T_h$ | eCK | GDH | 2 |
| He(2012),[11] | $5T_{mul} + T_{add} + T_{inv} + T_h$ | mBR | DCDH | 3 |
| He(2011),[12] | $5T_{mul} + 4T_{add} + 2T_h$ | mBR | CDH | 2 |
| He(2011),[19] | $5T_{mul} + 3T_{add} + 2T_h$ | eCK | GDH | 2 |
| Our Protocol | $4T_{mul} + 3T_{add} + 2T_h$ | eCK | GDH | 2 |

## 6. Conclusion

We proposed a bilinear pairing-free CTAKA protocol and proved its security in the eCK model.

In this protocol the number of scalar multiplication is 4.

This protocol has the best performance among related protocols.

**Acknowledgement** :Authors would like to thank anonymous reviewers' help and advice.